# THE AMAZING NORMAL FORCES

Horia I. Petrache
Department of Physics, Indiana University Purdue University Indianapolis
Indianapolis, IN 46202

November 9, 2012

**Abstract**

This manuscript is written for students in introductory physics classes to address some of the common difficulties and misconceptions of the normal force, especially the relationship between normal and friction forces. Accordingly, it is intentionally informal and conversational in tone to teach students how to build an intuition to complement mathematical formalism. This is accomplished by beginning with common and everyday experience and then guiding students toward two realizations: (*i*) That real objects are deformable even when deformations are not easily visible, and (*ii*) that the relation between friction and normal forces follows from the action-reaction principle. The traditional formulae under static and kinetic conditions are then analyzed to show that peculiarity of the normal-friction relationship follows readily from observations and knowledge of physics principles.

## 1. Normal forces: amazing or amusing?

Learning about normal forces can be a life changing event. In introductory physics, we accept and embrace these totally mysterious things. Suddenly, normal forces become a convenient answer to everything: they hold objects on floors, on walls, in elevators and even on ceilings. They lift heavy weights on platforms, let footballs bounce, basketball players jump, and as if this was not enough, they even tell friction what to do. (Ah, the amazing friction forces – yet another amazing story! [1]) Life before physics becomes inexplicable.

This article is about building an intuition about normal forces using the action-reaction law of mechanics and the fact that real objects are deformable. The thinking can start with looking around the room: there are chairs, tables, cabinets, and people, all sitting on the floor. Imagine how many normal forces are all around! Now, if you think about it, the floor must be very, very smart since it knows exactly where and when to apply what normal force. For example, if the teacher moves a chair and takes its place, the floor immediately brings in the correct normal force. Somehow, the floor can tell the teacher from other objects in the room. We conclude that we must find these marvelous people who build such smart floors as they must be the holders of the secret of the normal force.

But there is no mystery. Normal forces are *deformation forces*. They can also be called contact forces. Contact means deformation, and deformation gives rise to contact forces. Normal forces are called as such because they are perpendicular to the surface of contact. In mathematics, normal means perpendicular. We learn that friction forces are also contact forces but they are parallel to the surface rather than perpendicular. Confusion regarding normal and friction forces can arise in at least two occasions. First, we must accept that friction forces are proportional to normal forces. This proportionality appears intuitive based on everyday experience, but is it an experimental result or a fundamental physics law? How to think about the fact that two perpendicular forces are proportional to one another? Second, while for objects in motion we write $F_f = \mu_k F_N$ with a definite equal sign, for the static case we write the less convincing inequality, $F_f \leq \mu_s F_N$. So maybe friction is not really that proportional to the normal force after all. Or is it? How do I make sure and pass this class?! Perhaps I should just answer B to all questions about friction and normal forces.

Fortunately, there is no need to do that. As shown in the following, normal forces and their relationship with friction can be visualized in an intuitive way once we accept that real objects are deformable. We will first discuss objects on flat floors and then move to objects on inclines and see how the action-reaction principle comes to the rescue in each case. We will also see some interesting aspects of action-reaction forces for accelerating objects.

## 2. Normal forces on floors

Figure 1 shows an "artistic" view of a floor with two objects on it. In such an exaggerated vision, floors are deformable objects (more like bed mattresses) that give in under the weight of objects placed on them. Elastic floors balance the weight of objects by deformation forces as shown in the figure. Floor deformations are in proportion to the weights placed on them but depend greatly on the consistency of the floor material. On ceramic floors for example, deformations are not visible by eye because their "spring constants" $k$ are very large and therefore deformations $\Delta x$ are on molecular scale (see for example Ref [2] for elasticities at molecular level.)

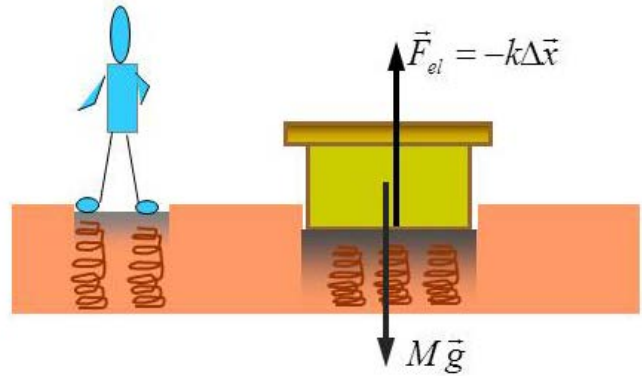

**Figure 1.** The mattress model of floors. Normal forces are deformation forces, although deformations must be grossly exaggerated to be shown in a picture like this. Deformations are harmonic in most "normal" situations. The arrows show the two forces acting on the desk: a gravitational force ($M\vec{g}$) and an elastic deformation force proportional to the "spring constant" of the floor ($k$) and the extent of deformation ($\Delta x$).

As depicted in Fig. 1, each object sitting on the floor acts with a force on the floor and the floor responds with a deformation force. The two forces are equal and opposite as stated by the action-reaction principle in mechanics (Newton's third law). The heavier the object, the larger the deformation of the floor and consequently the larger the normal force. We see that objects at equilibrium on flat floors present no difficulties once we account for deformability and forces associated with it. The situation then seems to be under control on flat floors. But can we apply the same reasoning to inclined surface where things are usually more complicated? Does the action-reaction principle continue to hold on inclined surfaces?

## 3. Normal forces on inclines

Incline situations as in Fig. 2 below can cause headaches for many reasons including proper use of trigonometry and the number of forces to be shown. The trigonometry problem requires practice and will not be addressed in detail here except to mention that forces are tilted away from the vertical by the same angle as the incline is titled away from the horizontal. As for the forces acting on the box, we learn that we must include the action of gravity ($F_{\text{grav}}$) and *two* *contact forces* due to the incline: a normal force $F_N$ and a friction force $F_f$. But why two reaction forces instead of one?!

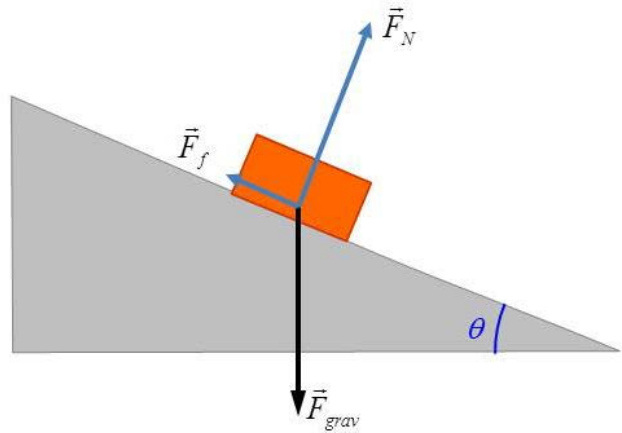

**Figure 2.** An object on an incline is shown to experience 3 forces: a gravitational force, a normal force, and a friction force.

According to the action-reaction principle, as the box presses its weight down on the incline, the incline should react with an equal and opposite force. This is shown in Fig. 3.

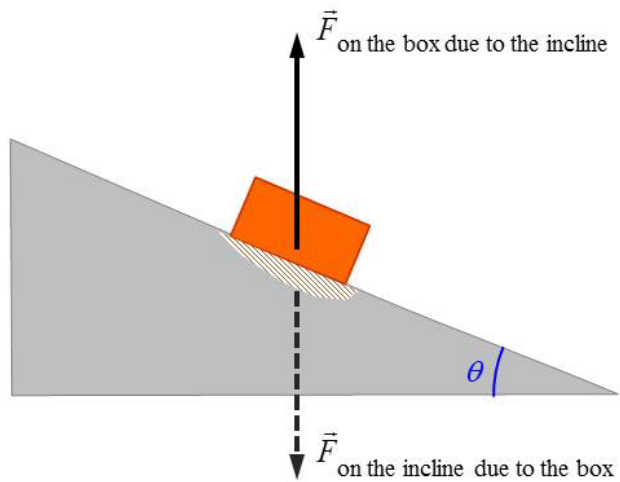

**Figure 3.** According to the action-reaction law, when the box acts with a force on the incline, the incline reacts with an equal and opposite force on the box. The reaction force is due to deformation at the contact area exaggeratedly shown here by the hashed area.

Figure 3 shows a pair of action-reaction forces: the box acts with force $F$ on the incline and the incline "responds" with an equal and opposite force on the box as stated by the action-reaction principle. What is usually not mentioned in textbooks is that we decide to decompose the reaction force into two components *for convenience*. One component is parallel to the surface of contact and the other is perpendicular to it. We call the former friction force and the second normal force. (The action-reaction principle does not restrict how we might like to decompose forces for our own benefit.) So one other mystery is easily solved: *the friction force and the normal force are related to one another precisely because they are the two components of a single reaction force*. This decomposition is shown in Fig. 4.

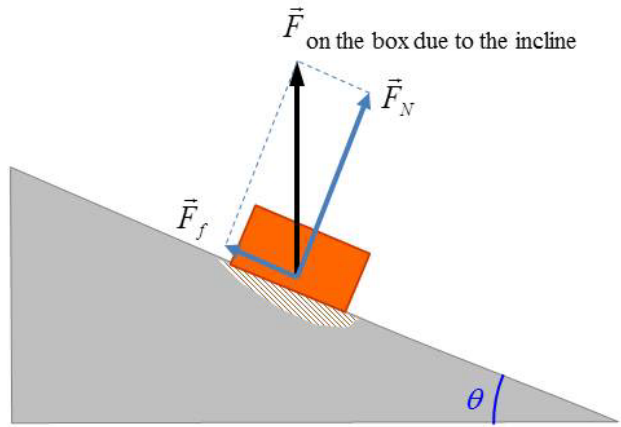

**Figure 4**. The reaction force experienced by the box due to the incline can be decomposed into two components: one parallel and the other perpendicular to the surface of contact. To complete the free body diagram of the box we need to add the gravitational force acting on the box.

The decomposition into a parallel and a perpendicular component is useful because in general we are interested in how the box moves (or not) along the contact surface. The parallel component (friction) affects the acceleration of the box along the incline, while the perpendicular component (normal force) tells us whether the box remains in contact with the surface or not. (No contact means zero normal force). However, we should not forget that being the two components of a single force, friction and normal force must be related to one another as we learn in class. Let us investigate this in more detail. From Fig. 4, we have:

$$F_f = F \sin \theta \qquad (1)$$
$$F_N = F \cos \theta, \qquad (2)$$

and assuming that we guessed correctly the *sin* and the *cos*, we end up with

$$F_f = F_N \tan \theta. \qquad (3)$$

So for a given tilt angle, the friction force is proportional to the normal force as we are told. This is true as long as the object does not move. However, we should expect that there is a maximum angle $\theta_{max}$ at which the object starts sliding down. The tangent of that maximum angle is called the static friction coefficient, $\mu_s$. Mathematically, $\mu_s = \tan \theta_{max}$. We then have

$$F_f = F_N \tan \theta \leq F_N \tan \theta_{max} = \mu_s F_N. \qquad (4)$$

*The inequality $F_f \leq \mu_s F_N$ is a consequence of the definition of the static coefficient $\mu_s$.* The dimensionless friction coefficient is a material parameter that is obtained by measurements, and it indicates the maximum tilt angle for a given pair of materials. Note that since the tangent function can have any positive value, so does the friction coefficient. Static friction coefficients therefore can be larger than 1, and in this case the friction force is larger than the normal force.

## 4. Normal and friction forces on objects in motion

This is a more complicated situation. Complications depend on whether the sliding object moves with constant speed or is accelerating. In any case, the reality is that what we learn in introductory physics about the kinetic friction coefficient $\mu_k$ is an approximation except for smooth objects at sufficiently small speeds. Writing $F_f = \mu_k F_N$ is inspired by $F_f = \mu_s F_N$ which holds for objects in equilibrium at maximum tilt. The coefficient $\mu_k$ is indeed smaller than $\mu_s$ as mentioned in textbooks but it can have a complicated dependence on speed and acceleration (see for example references [3] and [4]). However, in many situations, the approximation $F_f \cong \mu_k F_N$, which simply says that friction is proportional to the normal force, turns out to be pretty good as often verified in instructional labs (e.g. Ref. [5]).

Visualizing deformations explains why $\mu_k$ is smaller than $\mu_s$. As you might expect, deformations take time to develop. When an object is placed on a floor, it takes somewhere from microseconds to seconds for deformation to settle into equilibrium. On an ideal spring-like mattress, it can take longer or forever! In the case of an object moving over a surface, deformations are transient and might not get a chance to develop fully -- hence the reduced friction than in the static case. Note also that static friction can also depend on time. Objects that are kept in contact for a long period of time tend to "get stuck". This is due to molecular interactions and diffusion across the surface of contact – a process that is sped up by the presence of adhesives.

The typical textbook cartoon showing the molecular roughness is also correct in explaining the reduced friction for an object in motion, although it does not say why a moving object appears to hover over surface corrugations. A more complete description must account for molecular relaxation times in relation to the object's time spent in one place, although it is not easy to find a general rule (more discussion below). It is nevertheless expected and acceptable that the kinetic friction coefficients depend on speed. However this dependence makes things quite complicated for accelerated motion. In a typical problem, we are given a value for the kinetic friction coefficient and are supposed to calculate the acceleration of the object. And we can do that by using $\vec{F} = m\vec{a}$. But the acceleration that we calculate is not going to hold for too long. As the object accelerates, its speed changes and therefore the friction coefficient changes too. So now we have to solve the problem again with a new value for $\mu_k$! But the reason we do not worry about this complication is because in most table-top situations the whole motion (or experiment) is over before significant changes in speed can take place.

## 5. Normal forces on objects in motion *without friction*

We are now in serious trouble. In the static case in Fig. 3, we have a pair of action-reaction forces that are both vertical. The normal component of the force on the box is off the vertical as shown in Fig. 4 but that is OK because friction takes care of the other component making sure that the net reaction force is vertical. But if friction is absent, the net force on the box due to the incline cannot be vertical anymore which means that the force on the incline due to the box should be off the vertical as well in order to obey the action-reaction principle. To see that this is indeed so, let us look at the equations of motion for the general case in which friction is present:

$$mg\sin\theta - F_f = ma \tag{5}$$
$$mg\cos\theta - F_N = 0, \tag{6}$$

which can be written

$$F_f = mg\sin\theta - ma \tag{7}$$
$$F_N = mg\cos\theta. \tag{8}$$

This gives

$$F_f^2 + F_N^2 = (mg\sin\theta - ma)^2 + (mg\cos\theta)^2 \leq (mg)^2 \tag{9}$$

and

$$\frac{F_f}{F_N} = \frac{mg\sin\theta - ma}{mg\cos\theta} \leq \tan\theta. \tag{10}$$

These results say that the reaction force of the incline is less than *mg* and it is not vertical but tilted towards the normal to the surface. In the limit of zero friction, the reaction force is the normal force, as expected. This is illustrated in Fig. 5.

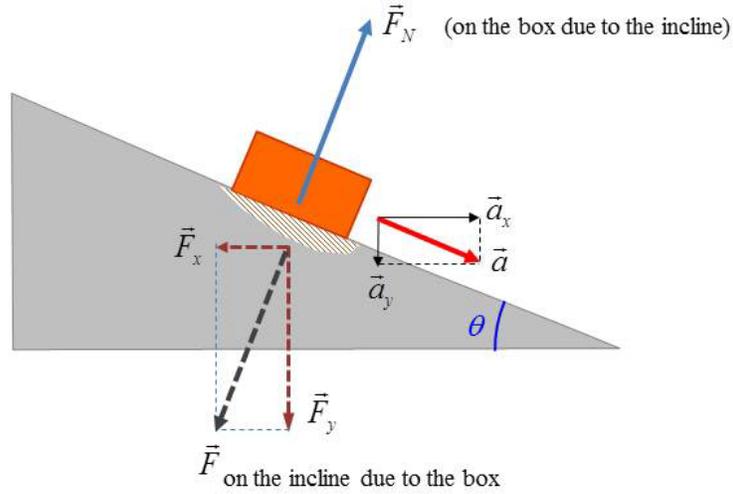

**Figure 5.** In the absence of friction, the box accelerates down the incline with acceleration $a = mg \sin \theta$. The action-reaction forces are perpendicular to the surface of contact.

For the non-friction case in Fig. 5, the force on the incline due to the box has to be equal and opposite to the normal force in order for the action-reaction principle to hold. Being tilted away from the vertical, the force on the incline due to the box has both a horizontal and a vertical component given by

$$F_x = F_N \sin \theta = mg \, \cos \theta \sin \theta \tag{11}$$
$$F_y = F_N \cos \theta = mg \, \cos \theta \cos \theta. \tag{12}$$

One way to "visualize" the horizontal component acting on the incline is to consider the case where there is no friction between the incline and the floor allowing the incline to move freely. In this case, because of conservation of momentum, the incline moves to the left as the box moves (accelerates) to the right. The horizontal component of the force reduces to zero for $\theta = 0$, as expected.

For the vertical component given by Eq. (12), think about the apparent weight of a falling object (like in problems with weights measured in elevators). The box accelerates downwards with acceleration $a_y = a \sin \theta = mg \sin \theta \sin \theta$. Its apparent weight then is $m(g - a_y) = mg(1 - sin^2\theta) = mg \, cos^2\theta$ which is precisely the force $F_y$ in Eq. (12). The box presses down on the incline with a force less than $mg$ because it falls with non-zero acceleration. The vertical component of the forced felt by the incline due to the box is less than $mg$ by a factor of $cos^2\theta$ and it becomes equal to $mg$ when $\theta = 0$, as expected.

For completeness, let us also write the expressions for $F_x$ and $F_y$ when friction is present. We have

$$F_x = F_N \sin \theta - F_f \cos \theta = mg \, (\cos \theta \sin \theta - \mu \, cos^2\theta) \tag{13}$$

$$F_y = F_N \cos\theta + F_f \sin\theta = mg\ (cos^2\theta + \mu \cos\theta \sin\theta)\ . \tag{14}$$

When $\mu = \tan\theta$, the vertical component $F_y$ becomes equal to $mg$ and the horizontal component $F_x$ becomes zero, recovering the static case shown in Fig. 2.

## 6. Concluding remarks and suggested reading

In conclusion, we have seen that proper consideration of the action-reaction law can answer many questions on normal and friction forces. Real objects are deformable whether deformations are easily visible or not, and taking deformations into account can make the physics or solid objects more realistic and possibly more revealing. Thinking in terms of rigid non-deformable idealized objects is less intuitive because the origin of normal forces is not immediately clear. Our focus here was on building intuition on normal forces and we addressed friction forces only to a very limited extent. We have shown that the proportionality relationship between friction and normal forces follows from the geometrical decomposition of a reaction force which is easily seen for the static case in Fig. 4. The case of accelerated motion is more complicated but it can be treated formally using Eqs. (7) and (8). We have

$$F_f = \frac{mg\ sin\theta - ma}{mg\ cos\theta} F_N = \left(\tan\theta - \frac{a}{g\ cos\theta}\right) F_N \equiv \mu_k F_N\ , \tag{15}$$

where we have identified the expression in parentheses as the kinetic friction coefficient, $\mu_k$. In this sense, friction and normal forces are proportional to one another but the proportionality factor $\mu_k$ is not necessarily a constant. According to Eq. (15), the object moves with constant speed ($a = 0$) for some particular value of $\theta$. As measured experimentally, this angle is smaller than the slipping angle $\theta_{max}$ for the static case. However, once the object is in motion and the tilt angle is varied, there is no guarantee that the friction coefficient stays the same. Even for smooth macroscopic objects on non-sticky surfaces (negligible adhesion) there are possible contributions from the geometry of surfaces. For example, friction can depend on the curvature of the object's front end -- think about the lifted fronts of skis and snowboards. The literature on friction coefficients is very rich especially in engineering and applied physics journals. A summary of results on static friction can be found in [6] and [7] and an interesting discussion of conditions for slipping on an incline can be found in [8]. Lastly, an instructive description of the atomic origin of friction forces can be found in [9] and references within.

**ACKNOWLEDGEMENTS**

Many thanks to Prof. Marvin Kemple for discussions, to Heather Stout and Andrew Seeran for editing of the manuscript, and special thanks to my students in General Physics classes for their challenging questions and for openly sharing their misunderstandings of normal and friction forces.

**REFERENCES**

[1] D. Grech, Z. Mazur, The amazing cases of motion with friction, Eur. J. Phys. **22**, 433- 440 (2001).
[2] T. Erber, Hooke's law and fatigue limits in micromechanics, Eur. J. Phys. **22**, 491-499, (2001).
[3] E. Rabinowicz, Stick and slip, Sci. Am. **194** (5), 109-119 (1956).
[4] H. L. Armstrong, How dry friction really behaves, Am. J. Phys. **53**, 910 – 911 (1985).
[5] L. M. Gratton, S. Defrancesco, A simple measurement of the sliding friction coefficient, Phys. Educ. **41**, 232- 235 (2006).
[6] V. Konecny, On the first law of friction, Am. J. Phys. **41**, 588-589 (1973).
[7] V. Konecny, On maximum force of static friction, Am. J. Phys. **41**, 733-734 (1973).
[8] W. M. Wehrbein, Frictional forces on an inclined plane, Am. J. Phys. **60**, 57 – 58 (1992).
[9] J. Ringlein, M. O. Robbins, Understanding and illustrating the atomic origins of friction, Am. J. Phys. **72**, 884- 891 (2004).